\documentstyle[aps,preprint,tighten,array,amsmath,epsfig,bm]{revtex}

\begin{document}

\title{Analytical calculations of four-neutrino oscillations in matter}

\author{
Yuki Kamo\footnote{Electronic address: {\tt kamo@sci.kumamoto-u.ac.jp}},
Satoshi Yajima, Yoji Higasida, Shin-Ichiro Kubota,
Shoshi Tokuo and Jun-Ichi Ichihara
}
\address{
Department of Physics, Kumamoto University, 
2-39-1 Kurokami, Kumamoto 860-8555, Japan}

\maketitle
\begin{abstract}
 We analytically derive transition probabilities 
 for four-neutrino oscillations in matter.
 The time evolution operator giving the neutrino oscillations is expressed by 
 the finite sum of terms up to the third power of the Hamiltonian
 in a matrix form, using the Cayley-Hamilton theorem.
 The result of computation for the probabilities 
 in some mass patterns tells us that
 it is realistically difficult to observe 
 the resonance between one of three active neutrinos and 
 the fourth (sterile) neutrino near the earth,
 even if the fourth neutrino exists.
\end{abstract}


\section{Introduction}
A neutrino oscillation is a transition among neutrino flavors.
Several types of the observations tell us that neutrino oscillations occur 
\cite{Ahmad01,Cleveland98,Hampel99,Fukuda96,Fukuda99,Scholberg99,%
Abdurashitov99,Athanassopoulos96}.
They are classified into the solar, atmospheric and LSND experiments.

The mass squared differences are the parameters 
showing the neutrino oscillations.
In order to explain three kinds of neutrino experiments within one framework,
three kinds of the mass squared differences are needed.
Therefore we consider the four-neutrino oscillation,
where the fourth neutrino doesn't have the weak interaction.
Three active neutrino flavors ($\nu_e,\nu_{\mu}, \nu_{\tau}$)
interact with leptons in the weak interaction.
So the fourth neutrino is called sterile neutrino ($\nu_s$).

The neutrino oscillation pattern in vacuum can get modified,
when the neutrinos pass through matter.
It is known as the Mikheyev-Smirnov-Wolfenstein (MSW) effect
\cite{MSW85},
which can be described by an effective Hamiltonian.
The interaction with the neutral currents occurs for three active neutrinos.
Thus, for the three active neutrinos,
one doesn't need to consider the interaction with the neutral currents
\cite{Bilenky98}.
But the sterile neutrino has neither the charged-
nor the neutral-current interactions.
It means that one needs to consider
the effect of the matter interacting with the sterile neutrino
and to introduce the $4 \times 4$ mixing matrix of four neutrinos
which is an extension of the $3 \times 3$ Maki-Nakagawa-Sakata (MNS)
matrix \cite{MNS62}.

Analytical calculations of active three-neutrino oscillations in matter
have been performed\cite{Ohlsson00}.
In this article, we derive analytically the transition probabilities 
for four-neutrino oscillations.
Our calculations include the effects of
the interaction with charged and neutral currents.

The outline of the article is as follows.
In Sec.~\ref{chap:formalism},
two kinds of the bases to express four-neutrino states are introduced.
These bases are connected by a mixing matrix.
To describe neutrino oscillations in matter,
the effective Hamiltonian with charged and neutral current is introduced.
In Sec.~\ref{chap:TransitionProbabilities},
we calculate the transition probabilities from the effective Hamiltonian. 
In order to derive the transition probabilities,
we make use of the Cayley-Hamilton theorem and
the formula for the root of biquadratic equation.
In Sec.~\ref{chap:4neutrinoosc},
the transition probabilities are concretely computed
in two cases of the four-neutrino oscillation schemes.
Finally, in Sec.~\ref{chap:discussion},
we discuss the effects of four-neutrino oscillations in matter.
    
\section{Formalism}\label{chap:formalism}
\subsection{Two bases and a mixing matrix}
Neutrinos are produced in flavor eigenstates
$ |{\nu_{\alpha}}\rangle (\alpha = e, \mu, \tau, s)$.
Between the source and the detector, the neutrinos evolve as 
mass eigenstates 
$ |{\nu_{a}}\rangle (a = 1, 2, 3, 4)$.
There are two kinds of eigenstates:
$ |{\nu_{\alpha}}\rangle$ and $|{\nu_{a}}\rangle$.
These eigenstates are defined by neutrino fields
$\nu_{\alpha}$ and $\nu_{a}$ corresponding to each eigenstate:
$\nu^{\dagger} |{0}\rangle \equiv |{\nu}\rangle$,
$|{\nu_{\alpha}}\rangle \equiv |{\alpha}\rangle$,
$|{\nu_{a}}\rangle \equiv |{a}\rangle$,
where a vacuum state is given by $|{0}\rangle$.
In the present analysis, 
we will use the plane wave approximation of the fields.
In this approximation, a neutrino flavor field $\nu_{\alpha}$ is
expressed by a linear combination of neutrino mass field $\nu_{a}$:
\begin{equation}
 \nu_{\alpha} = \sum _{a=1}^{4} U_{\alpha a} \nu_{a},
  \label{Eq:plane-wave-1}
\end{equation}
where $U$ is a 4 $\times$ 4 unitary matrix 
with the elements $U_{\alpha a}$.
If we write this relation in neutrino eigenstates, then
\begin{equation}
 |{\alpha}\rangle = \sum_{a=1}^{4} U_{\alpha a}^{*} |{a}\rangle.
\end{equation}

An arbitrary neutrino state $\psi$ is expressed in both the flavor and
mass bases as 
\begin{equation}
 \psi \equiv \sum_{\alpha = e, \mu, \tau, s} \psi_{\alpha} |{\alpha}\rangle
  =
  \sum_{\alpha = e, \mu, \tau, s} 
  \psi_{\alpha} \sum_{a=1}^{4} U_{\alpha a}^{*} |{a}\rangle
  =
  \sum_{a=1}^{4} \left(
                  \sum_{\alpha = e, \mu, \tau, s}\psi_{\alpha} U_{\alpha a}^{*}
                \right) |{a}\rangle
  =
  \sum_{a=1}^{4} \psi_{a}|{a}\rangle,
\end{equation}
where $\psi_{\alpha}$ and $\psi_{a}$ are 
the components of $\psi$ of the flavor eigenstate basis and
the mass eigenstate basis, respectively.
They are related to each other in the form with respect to
\begin{equation}
 \psi_{a} = \sum_{\alpha = e, \mu, \tau, s} U_{\alpha a}^{*} \psi_{\alpha}.
\end{equation}
If we define the matrix elements as
\begin{eqnarray}
 \psi_{f} &=& (\psi_{\alpha}) 
  = \left( 
     \begin{array}{c} 
      \psi_e \\
      \psi_{\mu} \\
      \psi_{\tau} \\
      \psi_{s}
     \end{array}
   \right),
\quad
 \psi_{m} = (\psi_{a}) 
 = \left( 
    \begin{array}{c}
     \psi_1 \\
     \psi_2 \\
     \psi_3 \\
     \psi_4
    \end{array}
  \right),
 \\
 U &=& (U_{\alpha a}) =
  \left(
   \begin{array}{cccc}
    U_{e1} & U_{e2} & U_{e3} & U_{e4} \\
    U_{\mu 1} & U_{\mu 2} & U_{\mu 3} & U_{\mu 4} \\
    U_{\tau 1} & U_{\tau 2} & U_{\tau 3} & U_{\tau 4} \\
    U_{s1} & U_{s2} & U_{s3} & U_{s4}
   \end{array}
 \right),
  \label{eq:unitary-matrix}
\end{eqnarray}
the relation between the flavor and the mass eigenstates are 
\begin{equation}
 \left(
  \begin{array}{c}
   \nu_e \\
   \nu_{\mu} \\
   \nu_{\tau} \\
   \nu_s 
  \end{array}
\right)
 =
 \left(
  \begin{array}{cccc}
   U_{e1} & U_{e2} & U_{e3} & {U_{e4}} \\
   U_{\mu 1} & U_{\mu 2} & U_{\mu 3} & {U_{\mu 4}} \\
   U_{\tau 1} & U_{\tau 2} & U_{\tau 3} & {U_{\tau 4}} \\
   {U_{s1}} & {U_{s2}} & {U_{s3}} & {U_{s4}}
  \end{array}
\right)
 \left(
  \begin{array}{c}
   \nu_1 \\
   \nu_2 \\
   \nu_3 \\
   \nu_4 
  \end{array}
\right).
\end{equation}
The unitary matrix $U$ is the mixing matrix of four neutrinos. 
There are 6 mixing angles and 3 mixing phases
as parameters of $U$, in the case of 4 neutrinos.
In this analysis, we ignore the CP violation 
by putting the mixing phases equal to zero.
Then, $U$ is the real orthogonal matrix\cite{Ohlsson00}.

A parameterization for 
$U=U(\theta_{12},\theta_{13},\theta_{14},\theta_{23},\theta_{24},\theta_{34})$
is given by
\begin{equation}
 U =U_{34}U_{24}U_{14}U_{23}U_{13}U_{12},
\end{equation}
where matrix elements are
\begin{eqnarray}
 &&{\left(
   U_{ij}
 \right)}_{ab}
 =
 \delta_{ab} + (C_{ij}-1)(\delta_{ia}\delta_{ib} + \delta_{ja}\delta_{jb})
 +S_{ij} (\delta_{ia}\delta_{jb} -\delta_{ja}\delta_{ib}),
 \\
 &&
  C_{ij} = \cos \theta_{ij},\quad S_{ij} = \sin \theta_{ij},
\end{eqnarray}
and the mixing angles
$\theta_{12},\theta_{13},\theta_{14},\theta_{23},\theta_{24},\theta_{34}$
\cite{Barger99,Barger00}.
By this definition, the mixing matrix becomes
\begin{eqnarray}
 &&
  U = \nonumber \\
 &&
 \left(
  \begin{array}{llll}
   \begin{array}{l}
    C_{12}C_{13}C_{14}\\
    \\
   \end{array}&
    \begin{array}{l}
     C_{13}C_{14}S_{12}\\
     \\
    \end{array}&
    \begin{array}{l}
     C_{14} S_{13}\\
     \\
    \end{array}&
    \begin{array}{l}
     S_{14}\\
     \\
    \end{array}\\
   \begin{array}{l}
    -C_{23}C_{24}S_{12} \\
    -C_{12}C_{24}S_{13}S_{23}\\
    -C_{12}C_{13}S_{14}S_{24}\\
    \\
   \end{array}&
   \begin{array}{l}
   C_{12}C_{23}C_{24}\\
   -C_{24}S_{12}S_{13} S_{23}\\
   -C_{13}S_{12}S_{14}S_{24}\\
    \\
    \end{array}&
    \begin{array}{l}
     C_{13}C_{24}S_{23}\\
    -S_{13} S_{14} S_{24}\\
     \\
    \end{array}&
   \begin{array}{l}
     C_{14} S_{24}\\
    \\
   \end{array} \\
   \begin{array}{l}
    -C_{12}C_{23}C_{34}S_{13}\\ 
    +C_{34}S_{12}S_{23}\\
    -C_{12}C_{13}C_{24}S_{14}S_{34} \\
    +C_{23}S_{12}S_{24}S_{34}\\
    +C_{12}S_{13}S_{23}S_{24}S_{34}\\
    \\
   \end{array}&
   \begin{array}{l}
    -C_{23}C_{34}S_{12}S_{13}\\
    -C_{12} C_{34} S_{23}\\
    -C_{13} C_{24} S_{12} S_{14} S_{34}\\
    -C_{12} C_{23} S_{24} S_{34}\\
    +S_{12} S_{13} S_{23} S_{24} S_{34}\\
     \\
   \end{array}&
   \begin{array}{l}
    C_{13} C_{23} C_{34}\\
    -C_{24} S_{13} S_{14} S_{34}\\
    -C_{13} S_{23} S_{24} S_{34}\\
    \\
   \end{array}&
   \begin{array}{l}
    C_{14} C_{24} S_{34}\\
    \\
   \end{array}\\
   \begin{array}{l}
    -C_{12} C_{13} C_{24} C_{34} S_{14} \\
    +C_{23} C_{34} S_{12} S_{24}\\
    +C_{12} C_{34} S_{13} S_{23} S_{24}\\
    +C_{12} C_{23} S_{13} S_{34}\\
    -S_{12} S_{23} S_{34}\\
    \\
   \end{array}&
   \begin{array}{l}
    -C_{13} C_{24} C_{34} S_{12} S_{14}\\
    -C_{12} C_{23} C_{34} S_{24}\\
    +C_{34} S_{12} S_{13} S_{23} S_{24}\\
    +C_{23} S_{12} S_{13} S_{34}\\
    +C_{12} S_{23} S_{34}\\
    \\
   \end{array}&
   \begin{array}{l}
    -C_{24} C_{34} S_{13} S_{14}\\
    -C_{13} C_{34} S_{23} S_{24}\\
    -C_{13} C_{23} S_{34}\\
   \end{array}&
   \begin{array}{l}
    C_{14} C_{24} C_{34}\\
   \end{array}\\
  \end{array}
 \right).
\label{eq:4x4mixingmatrix}
\end{eqnarray}
  
\subsection{Hamiltonian in matter}
In the mass eigenstate basis, the Hamiltonian ${\mathcal{H}}_0$ 
participating in the propagation of neutrinos in vacuum is given by 
\begin{equation}
 {\mathcal{H}}_0 = \left(
        \begin{array}{cccc}
         E_1 & 0 & 0 & 0 \\
         0 & E_2 & 0 & 0 \\
         0 & 0 & E_3 & 0 \\
         0 & 0 & 0 & E_4 \\
        \end{array}
      \right) ,
\end{equation}
where $E_a ~ (a=1,2,3,4)$ are the energies of 
the neutrino mass eigenstates $|{a}\rangle$ with mass 
$m_a$:
\begin{equation}
  E_a = \sqrt{{m_a}^2 + \bm{p}^2}.\label{Eq:on-shell}
\end{equation}
Here and hereafter 
we assume the momentum $\bm{p}$ to be the same for all mass eigenstates.

There are two kinds of the additional potentials for
describing the interactions between neutrinos and matter.
One is the interaction of the charged particles (electrons)
and its neutrino $\nu_e$:
       \begin{equation}
        V_e = \sqrt{2} G_F ~\text{diag}(N_e, 0, 0, 0).
       \end{equation}
The other is the interaction of the neutral particles(e.g. the neutron) 
and active neutrinos ($\nu_e,\nu_{\mu},\nu_{\tau}$):
       \begin{equation}
        V_{n0} = \sqrt{2} G_F ~
         \text{diag}(-\frac{1}{2}N_n, -\frac{1}{2}N_n, -\frac{1}{2}N_n, 0),\label{Eq:NC-interacton}
       \end{equation}
where $G_F$, $N_e$ and $N_n$ are 
the Fermi weak coupling constant, the electron number density
and the neutral particle number density, respectively.
Note that we assume the particle number densities to be
constant throughout the matter where the neutrinos are propagating.

The interaction term (\ref{Eq:NC-interacton}) 
can be separated into two parts as
\begin{eqnarray}
 V_{n0} &=& V_n + V^{\prime}, \\
 V_n &=& \sqrt{2} G_F ~ \text{diag}(0, 0, 0, +\frac{1}{2}N_n),\\
 V^{\prime} &=& \sqrt{2} G_F ~
  \text{diag}(-\frac{1}{2}N_n, -\frac{1}{2}N_n, -\frac{1}{2}N_n, -\frac{1}{
2}N_n).
\end{eqnarray}
These interaction terms are written by the flavor eigenstate basis.
Therefore the interaction terms in the flavor eigenstate basis must be
transformed into those in the mass eigenstate basis by the mixing matrix $U$. 
The interaction terms in the mass eigenstate basis are
\begin{eqnarray}
 U^{-1} V_e U
 &=& A_e \left(
        \begin{array}{cccc}
         U_{e1}^2 & U_{e1}U_{e2} & U_{e1}U_{e3} & U_{e1}U_{e4} \\
         U_{e2}U_{e1} & U_{e2}^2 & U_{e2}U_{e3} & U_{e2}U_{e4} \\
         U_{e3}U_{e1} & U_{e3}U_{e2} & U_{e3}^2 & U_{e3}U_{e4} \\
         U_{e4}U_{e1} & U_{e4}U_{e2} & U_{e4}U_{e3} & U_{e4}^2 \\
        \end{array} 
      \right),
 \\
 U^{-1} V_n U
 &=& A_n \left(
        \begin{array}{cccc}
         U_{s1}^2 & U_{s1}U_{s2} & U_{s1}U_{s3} & U_{s1}U_{s4} \\
         U_{s2}U_{s1} & U_{s2}^2 & U_{s2}U_{s3} & U_{s2}U_{s4} \\
         U_{s3}U_{s1} & U_{s3}U_{s2} & U_{s3}^2 & U_{s3}U_{s4} \\
         U_{s4}U_{s1} & U_{s4}U_{s2} & U_{s4}U_{s3} & U_{s4}^2 \\
        \end{array} 
      \right),
 \\
 U^{-1} V^{\prime} U &=& -A_n I, 
\end{eqnarray}
where $I$ is the $4 \times 4$ unit matrix, 
and the matter densities $A_e, A_n$ and $A$ are defined by
\begin{eqnarray}
  A_e &=& \sqrt{2} G_F N_e \equiv A, \label{eq:defA-e}\\
 A_n &=& \frac{1}{\sqrt{2}} G_F N_n = 
  \frac{1}{2} A \frac{N_n}{N_e}. \label{eq:defA-n}
\end{eqnarray}
Thus, the Hamiltonian in the case when the neutrinos propagate in matter is 
\begin{equation}
 {\mathcal{H}}_m = {\mathcal{H}}_0 + U^{-1}V_e U + U^{-1}V_n U - A_n I
  \label{eq:HamiltonianInMatter}.
\end{equation}

\section{Calculations of the neutrino transition probabilities}
\label{chap:TransitionProbabilities}
The transition probabilities are represented by the time evolution operator.
In the flavor state basis, the unitary transformation
from the initial state $\psi_f(t=0)$ to the final state $\psi_f(t)$ is 
given by the operator
\begin{equation}
 U_f(t) \equiv U_f(t, 0),
\end{equation}
where $U_f(t_2, t_1)$ is the time evolution operator
from time $t_1$ to $t_2$ in the flavor state basis.
The Hamiltonian ${\mathcal{H}}_{\textrm{flavor}}$ in the flavor state basis
is represented by using the mixing matrix $U$ 
and the Hamiltonian ${\mathcal{H}}_m$ in the mass state basis:
\begin{equation}
 {\mathcal{H}}_{\textrm{flavor}} = U {\mathcal{H}}_m U^{-1}.
\end{equation}
The Schr\"{o}dinger equation in the mass eigenstate basis is
\begin{equation}
 i \frac{d}{d t}\psi_m(t) = {\mathcal{H}}_m \psi_m(t).
\label{Eq:schrodinger-eq}
\end{equation}

Equation (\ref{Eq:schrodinger-eq}) has a solution
\begin{equation}
 \psi_m(t) = e^{-i {\mathcal{H}}_m t} \psi_m(0),
  \label{Eq:schrodinger-sol}
\end{equation}
where $e^{-i {\mathcal{H}}_m t}$ is the time evolution operator.
Inserting $t=L$ into (\ref{Eq:schrodinger-sol}),
the solution of the Schr\"{o}dinger equation (\ref{Eq:schrodinger-eq}) 
is 
\begin{equation}
 \psi_m(L) = \psi_m(t)\Big|_{t=L} = e^{-i {\mathcal{H}}_m L} \psi_m(0)
  \equiv U_m(L) \psi_m(0),
\end{equation}
where $L$ stands for the distance through which neutrinos run for the time $t$,
because the speed of neutrinos is almost equal to that of light.

The neutrino state $\psi_f(L)$ at $t=L$ in the flavor state basis is 
expressed as
\begin{equation}
  \psi_f(L) = U \psi_m(L) = U e^{-i {\mathcal{H}}_m L} \psi_m(0)
   = U e^{-i {\mathcal{H}}_m L} U^{-1} \psi_f(0) 
 \equiv U_f(L) \psi_f(0). 
\end{equation}
\subsection{Traceless matrix $T$}
In order to find the explicit form of the time evolution operator
$e^{-i {\mathcal{H}}_mt}$, which is the exponential of the matrix,
the Hamiltonian in the matrix form is 
separated into the diagonal and the traceless matrices.
The trace of the matrix ${\mathcal{H}}_m$ in (\ref{eq:HamiltonianInMatter}) is
\begin{equation}
 {\mathrm{tr}} {\mathcal{H}}_m = E_1 + E_2 + E_3 + E_4 + A_e -3 A_n,
\end{equation}
where we use the unitarity conditions, e.g.
${U_{e1}}^2 +{U_{e2}}^2 +{U_{e3}}^2 +{U_{e4}}^2 = 1$.

An arbitrary $N \times N$ matrix $M$ can always be written as
\begin{equation}
 M = M_0 + \frac{1}{N} ({\mathrm{tr}} M) I_N
  \label{eq:M-2-M01},
\end{equation}
where $M_0$ and $I_N$ are 
$N \times N$ traceless and unit matrices, respectively.
Note that ${\mathrm{tr}} M_0=0$.
Then the $4 \times 4$ matrix ${\mathcal{H}}_m$ can be written as
\begin{equation}
  {\mathcal{H}}_m = T + \frac{1}{4}({\mathrm{tr}} {\mathcal{H}}_m) I,
\end{equation}
where $I_4 = I$ and the matrix $T$ is traceless.
The matrix $T$ can be written by
\begin{eqnarray}
 T &=& {\mathcal{H}}_0 -\frac{1}{4}(E_1 + E_2 + E_3 + E_4)I
  + U^{-1} V_e U + U^{-1} V_n U 
  -\frac{1}{4}(A_e + A_n) I \nonumber \\
 &=& \frac{1}{4} 
  \left(
   \begin{array}{cccc}
    E_{1,234} & 0 & 0 & 0 \\
    0 & E_{2,134} & 0 & 0 \\
    0 & 0 & E_{3,124} & 0 \\
    0 & 0 & 0 & E_{4,123} \\
   \end{array}
 \right) 
  \nonumber \\
  && +
  A_e \left(
       \begin{array}{cccc}
        {U_{e1}}^2 -\frac{1}{4} & {U_{e1}}{U_{e2}} 
         & {U_{e1}}{U_{e3}} & {U_{e1}}{U_{e4}} \\
        {U_{e2}}{U_{e1}} & {U_{e2}}^2 -\frac{1}{4}
         & {U_{e2}}{U_{e3}} & {U_{e2}}{U_{e4}} \\
        {U_{e3}}{U_{e1}} & {U_{e3}}{U_{e2}} 
         & {U_{e3}}^2 -\frac{1}{4} & {U_{e3}}{U_{e4}}\\
        {U_{e4}}{U_{e1}} & {U_{e4}}{U_{e2}} 
         & {U_{e4}}{U_{e3}} & {U_{e4}}^2 -\frac{1}{4} \\
       \end{array}
     \right)
  \nonumber \\
  &&
 +
  A_n \left(
       \begin{array}{cccc}
        {U_{s1}}^2 -\frac{1}{4} & {U_{s1}}{U_{s2}} 
         & {U_{s1}}{U_{s3}} & {U_{s1}}{U_{s4}} \\
        {U_{s2}}{U_{s1}} & {U_{s2}}^2 -\frac{1}{4}
         & {U_{s2}}{U_{s3}} & {U_{s2}}{U_{s4}} \\
        {U_{s3}}{U_{s1}} & {U_{s3}}{U_{s2}} 
         & {U_{s3}}^2 -\frac{1}{4} & {U_{s3}}{U_{s4}}\\
        {U_{s4}}{U_{s1}} & {U_{s4}}{U_{s2}} 
         & {U_{s4}}{U_{s3}} & {U_{s4}}^2 -\frac{1}{4} \\
       \end{array}
     \right),
\end{eqnarray}
where $E_{ab}~ (a,b=1,2,3,4, a \neq b)$ and 
$E_{k,lmn} (k,l,m,n=1,2,3,4)$ are defined as
\begin{equation}
 E_{ab}  \equiv  E_a - E_b, \quad
  E_{k,lmn} \equiv E_{kl} + E_{km} + E_{kn},
 \nonumber
\end{equation}
respectively.
The energy differences $E_{ab}$ are not linearly independent,
since they obey the following relations:
\begin{eqnarray*}
 E_{ab} = - E_{ba},\quad 
 E_{12} + E_{23} + E_{31} = 0,\quad 
 E_{12} + E_{24} + E_{41} = 0,\quad 
 E_{13} + E_{34} + E_{41} = 0. 
\end{eqnarray*}
Thus, only three of the $E_{ab}$'s are linearly independent. 

Therefore the time evolution operator $e^{-i {\mathcal{H}}_mt}$
can be rewritten by the traceless matrix $T$:
\begin{equation}
 U_m(L) =
 e^{-i {\mathcal{H}}_m L} = \phi e^{-i T L},\label{Eq:time-evolution-op}
\end{equation}
where
 $\phi = e^{-{i}({\mathrm{tr}} {\mathcal{H}}_m) L/4}$
is a phase factor.

\subsection{The Cayley-Hamilton theorem}
In order to find the concrete form of the definite matrix $e^{-i TL}$, 
we use the Cayley-Hamilton theorem.
The exponential of the $4 \times 4$ matrix $T$ can be expressed by 
an infinite series
\begin{equation}
 e^{-i T L} = k_0 I + k_1 T + k_2 T^2 + k_3 T^3 + k_4 T^4 + \cdots
  \nonumber ,
\end{equation}
where $k_n=\frac{{(-iL)}^n}{n!} \quad (n=1,2,3,4,\cdots)$. 
The Cayley-Hamilton theorem implies that the eigenvalue $\lambda$ 
in the characteristic equation
\begin{equation}
 {\mathrm{det}} (T - \lambda I_4) =
  {\lambda}^4 + c_3 {\lambda}^3 + c_2 {\lambda}^2 + c_1 {\lambda} + c_0 = 0
  \label{Eq:characteristic-equation}
\end{equation}
of the matrix $T$ can be replaced with $T$ to give
\begin{equation}
{T}^4 + c_3 {T}^3 + c_2 {T}^2 + c_1 {T} + c_0 = 0 \label{Eq:characteristic-T}, 
\end{equation}
where $c_j$ $(j=0,1,2,3)$ are coefficients.
Using (\ref{Eq:characteristic-T}) repeatedly, 
the matrix $e^{-i TL}$ can be formally written 
in the form
\begin{equation}
 e^{-i T L} = a_0 I + a_1 T + a_2 T^2 + a_3 T^3
  \label{eq:equation4},
\end{equation}
where $a_j$ $(j=0,1,2,3)$ are coefficients which differ from
$k_n$ and $c_j$ in general.  

Because $T$ is the definite matrix, we need to find the coefficients $a_j$ 
explicitly in order to obtain the matrix $e^{-i TL}$.
If the characteristic equation (\ref{Eq:characteristic-equation})
has four solutions $\lambda_k$ $(k=1,2,3,4)$,
one can write the eigenvalues of $e^{-i TL}$ as
\begin{equation}
 e^{-i {\lambda_k} L} 
  = a_0 + a_1 {\lambda_k} + a_2 {\lambda_k}^2 + a_3 {\lambda_k}^3.
\label{Eq:matrix-eTL2lambda}
\end{equation}

By defining the following matrices,
\begin{eqnarray}
 \bm{e} = \left(
            \begin{array}{c}
             e^{-iL \lambda_1} \\
             e^{-iL \lambda_2} \\
             e^{-iL \lambda_3} \\
             e^{-iL \lambda_4}
            \end{array}
          \right), \quad
        \Lambda = \left(
                   \begin{array}{cccc}
                    1 & - i L {\lambda_1} & - L^2 {\lambda_1}^2  
                     & +i L^3 {\lambda_1}^3 \\
                    1 & - i L {\lambda_2} & - L^2 {\lambda_2}^2  
                     & +i L^3 {\lambda_2}^3 \\
                    1 & - i L {\lambda_3} & - L^2 {\lambda_3}^2  
                     & +i L^3 {\lambda_3}^3 \\
                    1 & - i L {\lambda_4} & - L^2 {\lambda_4}^2  
                     & +i L^3 {\lambda_4}^3
                   \end{array}
                 \right), \quad
        \bm{a} = \left(
                   \begin{array}{c}
                    a_0 \\
                    a_1 \\
                    a_2 \\
                    a_3 \\
                   \end{array}
                 \right),
        \label{eq:define-matrixform}
\end{eqnarray}
(\ref{Eq:matrix-eTL2lambda}) is written as the matrix form
 $\bm{e} = \Lambda \bm{a}$.
Then, one obtain the coefficient $\bm{a}$ by
\begin{equation}
 \bm{a}=\Lambda^{-1} \bm{e}.\label{eq:a-Lambda-e}
\end{equation}
Therefore, we should find the eigenvalues $\lambda_k$ of the matrix $T$ 
in order to know $\bm{a}$. 

\subsection{Characteristic equation of the matrix $T$}
In order to obtain the eigenvalues $\lambda_k$ of the matrix $T$,
one must solve the characteristic equation
\begin{equation}
 \left|
  \begin{array}{cccc}
   T_{11} - \lambda & T_{12} & T_{13} & T_{14} \\
   T_{21} & T_{22} - \lambda & T_{23} & T_{24} \\
   T_{31} & T_{32} & T_{33} - \lambda & T_{34} \\
   T_{41} & T_{42} & T_{43} & T_{44} - \lambda
  \end{array}
\right|
 = 
 {\lambda}^4 + c_3 {\lambda}^3 + c_2 {\lambda}^2 + c_1 {\lambda}+ c_0 
 = 0
  \label{eq:4d-eigen-eq}.
\end{equation}
The coefficient $c_0$ is the determinant of $T$,
$c_1$ and $c_2$ are expressed by the sum of cofactors of $T$,
and $c_3$ is given by the trace of $T$,
\begin{eqnarray*}
 c_3 &=& -(T_{11} + T_{22} + T_{33} + T_{44}) = - {\mathrm{tr}} T = 0, \\
\end{eqnarray*}
\begin{eqnarray*}
 c_2 &=& +T_{11}T_{22} + T_{11}T_{33} + T_{11}T_{44} 
  + T_{22}T_{33} + T_{22}T_{44} + T_{33}T_{44} \nonumber \\
 & & - T_{12}T_{21} - T_{13}T_{31} - T_{14}T_{41} 
  - T_{23}T_{32} - T_{24}T_{42} - T_{34}T_{43}, \\
\end{eqnarray*}
\begin{eqnarray*}
 c_1 &=& -T_{11}T_{22}T_{33} -T_{11}T_{22}T_{44}
  -T_{11}T_{33}T_{44} -T_{22}T_{33}T_{44} \nonumber \\
 && +T_{11}T_{23}T_{32} +T_{11}T_{24}T_{42} +T_{11}T_{34}T_{43}
  +T_{22}T_{13}T_{31} +T_{22}T_{14}T_{41} +T_{22}T_{34}T_{43} 
  \nonumber \\
 && +T_{33}T_{12}T_{21} +T_{33}T_{14}T_{41} +T_{33}T_{24}T_{42}
  +T_{44}T_{12}T_{21} +T_{44}T_{13}T_{31} +T_{44}T_{23}T_{32} 
  \nonumber \\
 && -T_{13}T_{34}T_{41} -T_{14}T_{43}T_{31}
  -T_{12}T_{24}T_{41} -T_{14}T_{42}T_{21} 
  \nonumber \\
 &&
  -T_{12}T_{23}T_{31} -T_{13}T_{32}T_{21}
  -T_{23}T_{34}T_{42} -T_{24}T_{43}T_{32} 
  ,\\
\end{eqnarray*}
\begin{eqnarray*}
 c_0 &=& T_{11}T_{22}T_{33}T_{44} 
 \nonumber \\
 && -T_{11}T_{22}T_{34}T_{43} -T_{11}T_{33}T_{24}T_{42}
  -T_{11}T_{44}T_{23}T_{32} 
  \nonumber \\ %
 && %
  -T_{22}T_{33}T_{14}T_{41}
  -T_{22}T_{44}T_{13}T_{31} -T_{33}T_{44}T_{12}T_{21}
  \nonumber \\
 &&
 +T_{11}T_{23}T_{34}T_{42} +T_{11}T_{24}T_{43}T_{32}
 +T_{22}T_{13}T_{34}T_{41} +T_{22}T_{14}T_{43}T_{31}
  \nonumber \\
 &&
 +T_{33}T_{12}T_{24}T_{41} +T_{33}T_{14}T_{42}T_{21}
 +T_{44}T_{12}T_{23}T_{31} +T_{44}T_{13}T_{32}T_{21}
  \nonumber \\
 &&
 -T_{12}T_{23}T_{34}T_{41} -T_{12}T_{24}T_{43}T_{31}
 -T_{13}T_{32}T_{24}T_{41} 
  \nonumber \\ %
 && %
 -T_{13}T_{34}T_{42}T_{21}
 -T_{14}T_{42}T_{23}T_{31} -T_{14}T_{43}T_{32}T_{21}
  \nonumber \\
 &&
  +T_{12}T_{21}T_{34}T_{43}
 +T_{13}T_{31}T_{24}T_{42} +T_{14}T_{41}T_{23}T_{32}.
\end{eqnarray*}

The four roots of the biquadratic equation (\ref{eq:4d-eigen-eq}) are given 
from the solutions of the two quadratic equations \cite{Spiegel68}
\begin{equation}
 X^2 \pm \sqrt{t_0 -c_2} X + \frac{t_0}{2} 
  + \sqrt{{\frac{t_0}{2}}^2 -c_0} = 0,
\end{equation}
where $t_0$ is one of real roots of the cubic equation
\begin{equation}
 t^3 - c_2 t^2 -4 c_0 t +4 c_0 c_2 -{c_1}^2 = 0.
\end{equation}
Note that $c_3= -{\mathrm{tr}}T =0$ due to the definition of $T$.

\subsection{Calculation of time evolution operator}
From (\ref{Eq:time-evolution-op}) and (\ref{eq:equation4}), %
the time evolution operator is written as
\begin{eqnarray}
 U_{m}(L) &=& e^{-i{\mathcal{H}}_m L} = \phi e^{-i TL} \nonumber \\
  &=& \phi \left[
            a_0 I + (-iLT) a_1 - L^2 T^2 a_2 + i L^3 T^3 a_3
          \right].\label{Eq:time-evolution-op2}
\end{eqnarray}
The coefficients $a_j$'s are given by (\ref{eq:a-Lambda-e}) as follows,
\begin{eqnarray*}
 a_0 &=&
  -
  \frac{\lambda_2 \lambda_3 \lambda_4}
  {\lambda_{12}\lambda_{13}\lambda_{14}}
  e^{-iL \lambda_1}
  -
  \frac{\lambda_1 \lambda_3 \lambda_4}
  {\lambda_{21}\lambda_{23}\lambda_{24}}
  e^{-iL \lambda_2}
  -
  \frac{\lambda_1 \lambda_2 \lambda_4}
  {\lambda_{31}\lambda_{32}\lambda_{34}}
  e^{-iL \lambda_3}
  -
  \frac{\lambda_1 \lambda_2 \lambda_3}
  {\lambda_{41}\lambda_{42}\lambda_{43}}
  e^{-iL \lambda_4}
  , \\ 
 a_1 &=&
 \frac{i}{L}
 \left(
  \frac{\lambda_2 \lambda_3 + \lambda_2 \lambda_4 + \lambda_3 \lambda_4}
  {\lambda_{12}\lambda_{13}\lambda_{14}}
  e^{-iL \lambda_1}
  +
  \frac{\lambda_1 \lambda_3 + \lambda_1 \lambda_4 + \lambda_3 \lambda_4}
  {\lambda_{21}\lambda_{23}\lambda_{24}}
  e^{-iL \lambda_2}
  \right.
  \nonumber \\ &&
  \left.
  +
  \frac{\lambda_1 \lambda_2 + \lambda_1 \lambda_4 + \lambda_2 \lambda_4}
  {\lambda_{31}\lambda_{32}\lambda_{34}}
  e^{-iL \lambda_3}
  +
  \frac{\lambda_1 \lambda_2 + \lambda_1 \lambda_3 + \lambda_2 \lambda_3}
  {\lambda_{41}\lambda_{42}\lambda_{43}}
  e^{-iL \lambda_4}
  \right)
  , \\ 
 a_2 &=&
  \frac{1}{L^2}
  \left(
  \frac{\lambda_2 + \lambda_3 + \lambda_4}
  {\lambda_{12}\lambda_{13}\lambda_{14}}
  e^{-iL \lambda_1}
  +
  \frac{\lambda_1 + \lambda_3 + \lambda_4}
  {\lambda_{21}\lambda_{23}\lambda_{24}}
  e^{-iL \lambda_2}
  \right.
  \nonumber \\ && \left.
  +
  \frac{\lambda_1 + \lambda_2 + \lambda_4}
  {\lambda_{31}\lambda_{32}\lambda_{34}}
  e^{-iL \lambda_3}
  +
  \frac{\lambda_1 + \lambda_2 + \lambda_3}
  {\lambda_{41}\lambda_{42}\lambda_{43}}
  e^{-iL \lambda_4}
  \right)
  , \\ 
 a_3 &=&
  \frac{-i}{L^3}
  \left(
  \frac{1}
  {\lambda_{12}\lambda_{13}\lambda_{14}}
  e^{-iL \lambda_1}
  +
  \frac{1}
  {\lambda_{21}\lambda_{23}\lambda_{24}}
  e^{-iL \lambda_2}
  +
  \frac{1}
  {\lambda_{31}\lambda_{32}\lambda_{34}}
  e^{-iL \lambda_3}
  +
  \frac{1}
  {\lambda_{41}\lambda_{42}\lambda_{43}}
  e^{-iL \lambda_4}
  \right),
\end{eqnarray*}
where $\lambda_{ab} \equiv \lambda_a - \lambda_b$.
Inserting these results into (\ref{Eq:time-evolution-op2}),
one can find the time evolution operator in terms of $\lambda_a$'s.
For example, the term containing $e^{-iL \lambda_1}$ is 
\begin{equation}
  -
  \frac{
   \lambda_2 \lambda_3 \lambda_4 I
   -
   (\lambda_2 \lambda_3 + \lambda_2 \lambda_4 + \lambda_3 \lambda_4) T
   +
   (\lambda_2 + \lambda_3 + \lambda_4) T^2
   -
   T^3
  }
  {{\lambda_1}^3 -(\lambda_2 + \lambda_3 + \lambda_4) {\lambda_1}^2
  -
  (\lambda_2 \lambda_3 + \lambda_2 \lambda_4 + \lambda_3 \lambda_4) \lambda_1
  -
  \lambda_2 \lambda_3 \lambda_4
  }
  e^{-iL \lambda_1} \label{Eq:calc-time-evolution}.  
\end{equation}
Using the relations of 
the coefficients and solutions for the biquadratic equations,
\begin{eqnarray}
 \lambda_1 + \lambda_2 + \lambda_3 + \lambda_4 &=& - c_3 = 0,\\
 \lambda_1 \lambda_2 + \lambda_2 \lambda_3 + \lambda_3 \lambda_4 +
 \lambda_4 \lambda_1 + \lambda_1 \lambda_3 + \lambda_2 \lambda_4
 &=& c_2 ,\\
 \lambda_1 \lambda_2 \lambda_3 + \lambda_2 \lambda_3 \lambda_4 + 
 \lambda_1 \lambda_2 \lambda_4 + \lambda_1 \lambda_3 \lambda_4
 &=& -c_1 ,\\
 \lambda_1 \lambda_2 \lambda_3 \lambda_4 &=& c_0 ,
\end{eqnarray}
(\ref{Eq:calc-time-evolution}) can be written as
\begin{equation}
 \frac{(c_1 + c_2 \lambda_1 + {\lambda_1}^3) I 
  + (c_2 + {\lambda_1}^2)T + {\lambda_1} T^2 + T^3}
  {4 {\lambda_1}^3 + c_1 + 2 c_2 \lambda_1}
  e^{-iL \lambda_1}.
\end{equation}
Therefore, the matrix $e^{-iTL}$ is given by
\begin{eqnarray}
 e^{-iTL} &=& 
  \sum_{a=1}^{4} B_a e^{-iL \lambda_a}, \\
 B_a &\equiv& 
 \frac{(c_1 + c_2 \lambda_a + {\lambda_a}^3) I 
  + (c_2 + {\lambda_a}^2)T + {\lambda_a} T^2 + T^3}
  {4 {\lambda_a}^3 + c_1 + 2 c_2 \lambda_a}.
\end{eqnarray}
The time evolution operator in the mass eigenstate 
is derived by
\begin{equation}
 U_m(L) = e^{-i {\mathcal{H}}_m L} =
  \phi   \sum_{a=1}^{4} B_a  e^{-iL \lambda_a}.
\end{equation}
Using the mixing matrix $U$,
the time evolution operator in the flavor eigenstate is given as
\begin{eqnarray}
  U_f(L) &=& e^{-i {\mathcal{H}}_f L} 
   = U e^{-i {\mathcal{H}}_m L} U^{-1}
  =
  \phi   \sum_{a=1}^{4} \tilde{B_a} e^{-iL \lambda_a}
  \label{eq:flavor-t-evol}
  ,\\
 \tilde{B_a} &\equiv& U {B_a} U^{-1}
  =  \frac{(c_1 + c_2 \lambda_a + {\lambda_a}^3) I 
  + (c_2 + {\lambda_a}^2)\tilde{T} 
  + {\lambda_a} \tilde{T}^2 + \tilde{T}^3}
  {4 {\lambda_a}^3 + c_1 + 2 c_2 \lambda_a},
\end{eqnarray}
where $\tilde{T} \equiv U T U^{-1}$.

\subsection{Transition probabilities in matter}
A probability amplitude is defined as 
\begin{equation}
 A_{\alpha \beta} \equiv \langle{\beta}|U_f(L)|{\alpha}\rangle
  ,\quad\alpha, \beta = e, \mu, \tau, s.
  \label{eq:probability-amplitude}
\end{equation}
Inserting (\ref{eq:flavor-t-evol}) 
into (\ref{eq:probability-amplitude}) the probability amplitude becomes
\begin{eqnarray}
 A_{\alpha \beta} &=&
  \phi   \sum_{a=1}^{4} \tilde{(B_a)}_{\alpha \beta} 
  e^{-iL \lambda_a}
  ,\\
 \tilde{(B_a)}_{\alpha \beta} &=& 
  \frac{(c_1 + c_2 \lambda_a + {\lambda_a}^3) \delta_{\alpha\beta} 
  + (c_2 + {\lambda_a}^2){\tilde{T}}_{\alpha\beta} 
  + {\lambda_a}^2 {(\tilde{T}^2)}_{\alpha\beta}
 + {(\tilde{T}^3)}_{\alpha\beta}}
  {4 {\lambda_a}^3 + c_1 + 2 c_2 \lambda_a}
  \label{eq:ProbabilityAmplitudes},
\end{eqnarray}
where 
\begin{equation}
 \langle{\alpha}|I |{\beta}\rangle = \delta_{\alpha\beta},\quad
 \langle{\alpha}|\tilde{T} |{\beta}\rangle = \tilde{T}_{\alpha\beta},\quad
 \langle{\alpha}|{\tilde{T}}^2 |{\beta}\rangle = 
 {({\tilde{T}}^2)}_{\alpha\beta},
 \quad
 \langle{\alpha}|{\tilde{T}}^3 |{\beta}\rangle =
 {({\tilde{T}}^3)}_{\alpha\beta}
 \nonumber.
\end{equation}

Here $\delta_{\alpha\beta}$, $\tilde{T}_{\alpha\beta}$,
${({\tilde{T}}^2)}_{\alpha\beta}$ and 
${({\tilde{T}}^3)}_{\alpha\beta}$ are all symmetric. 
The probability of transition
from the neutrino flavor $\alpha$ to the neutrino flavor $\beta$ is
defined by
\begin{equation}
 P_{\alpha\beta} \equiv
  \left|
   A_{\alpha\beta}
 \right|^2
  =
  {A_{\alpha\beta}}^{*}A_{\alpha\beta}.
\end{equation}
Using the definition of the probability amplitude
(\ref{eq:probability-amplitude}),
one finds 
\begin{equation}
 P_{\alpha\beta}
  = \sum_{a=1}^{4}\sum_{b=1}^{4}
  \tilde{(B_a)}_{\alpha \beta} \tilde{(B_b)}_{\alpha \beta}
  e^{-iL(\lambda_a - \lambda_b)}
  \label{eq:calc-probability-amplitude},
\end{equation}
where the symmetry of $T$ leads $P_{\alpha\beta} = P_{\beta\alpha}$.

The probabilities for the oscillations in vacuum are given by 
setting $A_e=A_n=0$ in the definition of the probability amplitude.
From (\ref{eq:HamiltonianInMatter}), one can find
 \begin{eqnarray*}
  \langle{b}|U_m(L) |{a}\rangle &=& 
   \langle{b}| e^{-i {\mathcal{H}}_m L} |{a}\rangle
  = \langle{b}| e^{-i {\mathcal{H}}_0 L} |{a}\rangle
  = e^{-i E_a L} \delta_{ab} 
  ,\quad a,b= 1,2,3,4,
 \end{eqnarray*}
where ${\mathcal{H}}_m \big|_{A_e=0, A_n=0} = {\mathcal{H}}_0$.
Setting the following parameters
\begin{equation}
 x_{ab} \equiv \frac{1}{2}E_{ab}L, \quad
 \delta_{\alpha \beta} 
 = \sum_{a}{U_{\alpha a}}^2{U_{\beta a}}^2
  +2\sum_{a < b}
  U_{\alpha a}U_{\beta a}U_{\alpha b}U_{\beta b},
\end{equation}
the probability amplitude $A_{\alpha\beta}$ and
the probability $P_{\alpha\beta}$ of transition are given by
\begin{eqnarray}
 A_{\alpha\beta}
  &=& \langle{\beta}| U_f(L) |{\alpha}\rangle
  = \langle{\beta}| U e^{-i {\mathcal{H}}_m L} U^{-1} |{\alpha}\rangle
 = \sum_{a=1}^{4} U_{\alpha a} U_{\beta a} e^{-i E_a L}, 
 \\
 P_{\alpha\beta}
 &=& \delta_{\alpha\beta}
  - 4 \sum_{a < b} U_{\alpha a}U_{\beta a}U_{\alpha b}U_{\beta b}
  \sin^2 x_{ab} \label{eq:Pab-vacuum}.
\end{eqnarray}

The expression (\ref{eq:flavor-t-evol}) gives the relations at $L=0$
\[
 \phi = e^{-iL {\mathrm{tr}} {\mathcal{H}}_m/4} \big|_{L=0} =1,
  \quad
 I = U U^{-1} = \sum_{a=1}^{4} \tilde{B_a},
 \quad
 \delta_{\alpha\beta} = \langle{\alpha}|I |{\beta}\rangle
 = \sum_{a=1}^{4} \tilde{(B_a)}_{\alpha \beta}.
 \nonumber
\]
Then, we can rewrite the probabilities (\ref{eq:calc-probability-amplitude})
 for oscillations in matter 
in a form analogous to (\ref{eq:Pab-vacuum}) in the vacuum,
\begin{equation}
 P_{\alpha\beta} = \delta_{\alpha\beta}
  - 4 
  \underset{a < b}{
  \sum_{\scriptstyle a=1}^{4}
  \sum_{\scriptstyle b=1 
   }^{4} 
   }
   \tilde{(B_a)}_{\alpha \beta} \tilde{(B_b)}_{\alpha \beta}
   \sin^2 \tilde{x}_{ab} ,
\label{Eq:probability-of-transition}
\end{equation}
where \begin{equation}
 \tilde{x}_{ab} = \frac{L}{2}(\lambda_a -\lambda_b).
\end{equation}
If one set $\alpha \neq \beta$, then it is written as 
\begin{eqnarray}
 P_{\alpha\beta}(\alpha \neq \beta) &=& 
  - 4 
  \underset{a < b}{
  \sum_{\scriptstyle a=1}^{4}
  \sum_{\scriptstyle b=1
  }^{4}
  }
  \tilde{(C_a)}_{\alpha \beta} \tilde{(C_b)}_{\alpha \beta}
  \sin^2 \tilde{x_{ab}} 
  ,\\
 \tilde{(C_a)}_{\alpha \beta} &\equiv& 
 \frac{ (c_2 + {\lambda_a}^2)\tilde{T}_{\alpha\beta}
 + {\lambda_a} {(\tilde{T}^2)}_{\alpha\beta}
  + {(\tilde{T}^3)}_{\alpha\beta}}
  {4 {\lambda_a}^3 + c_1 + 2 c_2 \lambda_a}.
\end{eqnarray}

From unitarity, (\ref{Eq:probability-of-transition}) gives the relations
\begin{eqnarray}
 P_{ee} + P_{e\mu} + P_{e\tau} + P_{es} &=& 1,\\
 P_{\mu e} + P_{\mu\mu} + P_{\mu\tau} + P_{\mu s} &=& 1,\\
 P_{\tau e} + P_{\tau\mu} + P_{\tau\tau} + P_{\tau s} &=& 1,\\
 P_{se} + P_{s\mu} + P_{s\tau} + P_{ss} &=& 1,
\end{eqnarray}
where
\begin{equation}
 P_{e\mu} = P_{\mu e},\quad  P_{e\tau} = P_{\tau e},\quad 
  P_{\mu\tau} = P_{\tau\mu},\quad P_{es} = P_{se},\quad
  P_{\mu s} = P_{s \mu},\quad P_{\tau s}=P_{s \tau}.
\end{equation}
Hence, there are only 6 independent transition probabilities.

\section{Four-neutrino oscillations in matter}\label{chap:4neutrinoosc}
We apply some results obtained in the previous section
to the four-neutrino oscillation models.
The probability (\ref{Eq:probability-of-transition}) 
for transition of four-neutrino oscillation in matter 
contains the neutrino energy differences
$E_{ab} = E_a - E_b$ $(a,b=1,2,3,4) $ in vacuum.
This energy differences are approximately given by the matter mass differences
$\Delta {m_{ab}^2}={m_{a}}^2-{m_{b}}^2$,
which are well-known quantities in various kinds of the neutrino experiments.
From the on-shell condition (\ref{Eq:on-shell}),
\begin{equation}
 \Delta {m_{ab}}^2 = {m_a}^2 - {m_b}^2
 = {E_a}^2 - {E_b}^2
 = (E_a - E_b)(E_a + E_b)
 = 2 E_{ab} \frac{E_a + E_b}{2}.
\end{equation}
Now, we assume the average $E$ of the neutrino energies 
is about $10 \mathrm{GeV}$ :
\begin{equation}
 E = \frac{E_1 + E_2 + E_3 +E_4}{4} \sim 10 \text{GeV}.
\end{equation}
One defines the difference $\delta$ between two kinds of the averages 
of two neutrino energies,
e.g., the average $\frac{E_1 + E_2}{2}$ of the neutrino \#1 and \#2 and
the average $\frac{E_3 + E_4}{2}$ of the neutrino \#3 and \#4.
We assume that the difference 
$\delta = \frac{E_3 + E_4}{2} -\frac{E_1 +E_2}{2}$ is
about $ 1\mathrm{eV}$ 
which is estimated from the result of the LSND experiment.
Then, 
\[
 E = \frac{\frac{E_1 + E_2}{2} + \frac{E_3 +E_4}{2}}{2}
 = \left( \frac{E_1 + E_2}{2} \right) +\frac{\delta}{2} \nonumber.
\]
It is found for $\delta \ll E$ that
\begin{equation}
 E_{ab} \simeq \frac{\Delta {m_{ab}}^2}{2E}.
\end{equation}

\begin{figure}[htb]
\begin{center}
 \begin{tabular}{cc}
   \includegraphics[height=4cm,clip]{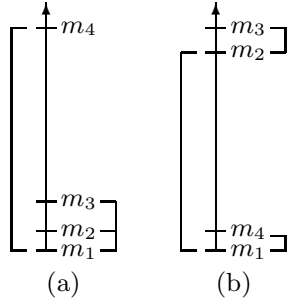}
 \end{tabular}
\end{center}
  \caption{Two of mass patterns of four neutrino schemes.
 (a):(3+1)-scheme, (b):(2+2)-scheme.}
  \label{fig:massscheme}
\end{figure}

In four-neutrino oscillation analysis,
there are three kinds of the mass squared differences.
They are used as the parameters in 
the solar and atmospheric oscillations and the LSND experiment,
which are represented as
$\Delta m_{\mathrm{solar}}^2$, $\Delta m_{\mathrm{atm}}^2$ and 
$\Delta m_{\mathrm{LSND}}^2$,
respectively.
Using these mass squared differences,
one can consider several distinct types of mass patterns
\cite{Bilenky98}.
They are classified into the so-called (3+1)-scheme and (2+2)-scheme.
We concentrate the discussion on two of the several mass patterns
in Fig.~\ref{fig:massscheme}.
The phenomenology and the mixing matrix depend on 
the type of the mass schemes.

\subsection{(3+1)-scheme}
We assume the mass patterns shown in Fig.~\ref{fig:massscheme}(a),
in which there are three close masses and one distinct mass.
Let $m_4$ and $\Delta m_{43}^2$ be the distinct mass
and the largest mass squared difference, respectively.

First, three kinds of the neutrino mass squared difference are put as follows
\cite{Ohlsson00,Barger00},
\begin{eqnarray}
 \Delta m_{21}^2 &=& \Delta m^2_{\mathrm{solar}}
  \simeq 10^{-4} \mathrm{eV^2}, \\
 \Delta m_{32}^2 &=& \Delta m^2_{\mathrm{atm}} 
  \simeq 10^{-3} \mathrm{eV^2}, \\
 \Delta m_{41}^2 &=& \Delta m^2_{\mathrm{LSND}}
  \simeq 1 \mathrm{eV^2}.
\end{eqnarray}
Then, the energy differences $E_{ab}$'s are expressed by
\begin{eqnarray}
 E_{21} \simeq \frac{\Delta m_{21}^2}{2E},\quad
 E_{32} \simeq \frac{\Delta m_{32}^2}{2E},\quad
 E_{41} \simeq \frac{\Delta m_{41}^2}{2E},\quad
 \nonumber \\
 E_{31} = E_{32} + E_{21},\quad
 E_{42} = E_{41} - E_{21},\quad
 E_{43} = E_{41} - E_{31},\quad
\end{eqnarray}
where we suppose that the average $E$ of the neutrino energies 
is $10 \mathrm{GeV}$.

Next, we consider the approximate mixing matrix for the (3+1)-scheme
\cite{Barger00}:
\begin{equation}
 \left(
 \begin{array}{c}
  \nu_e \\
  \nu_{\mu} \\
  \nu_{\tau} \\
  \nu_s 
 \end{array}
 \right)
 =
\left(
\begin{array}{cccc}
 \frac{1}{\sqrt{2}}\cos \epsilon & \frac{1}{\sqrt{2}}\cos \epsilon &
  0 & \sin \epsilon \\
 -\frac{1}{2} & \frac{1}{2} &
  \frac{1}{\sqrt{2}} & 0 \\
 \frac{1}{2} & -\frac{1}{2} &
  \frac{1}{\sqrt{2}} & 0\\
 \frac{-1}{\sqrt{2}}\sin \epsilon & \frac{-1}{\sqrt{2}}\sin \epsilon&
  0 & \cos \epsilon\\
\end{array} 
\right)
 \left(
 \begin{array}{c}
  \nu_1 \\
  \nu_2 \\
  \nu_3 \\
  \nu_4 
 \end{array}
 \right),
 \label{Eq:31mixingmatrix}
\end{equation}
where $\epsilon$ is small: $0 \leq \epsilon \leq 0.1 $.
Here the $3 \times 3$ sub-matrix that describes the mixing of 
the three active neutrinos has the bimaximal form.
The mixing matrix of (\ref{Eq:31mixingmatrix}) 
is given from (\ref{eq:4x4mixingmatrix}) by taking 
\begin{equation}
 \theta_{12} = \frac{\pi}{4},\quad \theta_{23} = \frac{\pi}{4},\quad
 \theta_{13} = 0,\quad \theta_{14} = \epsilon,
\quad \theta_{24} = \delta = 0,\quad \theta_{34} = 0.
\end{equation}

\begin{figure}[htbp]
 \begin{center}
  \begin{tabular}{c}
   \includegraphics[height=6cm,clip]{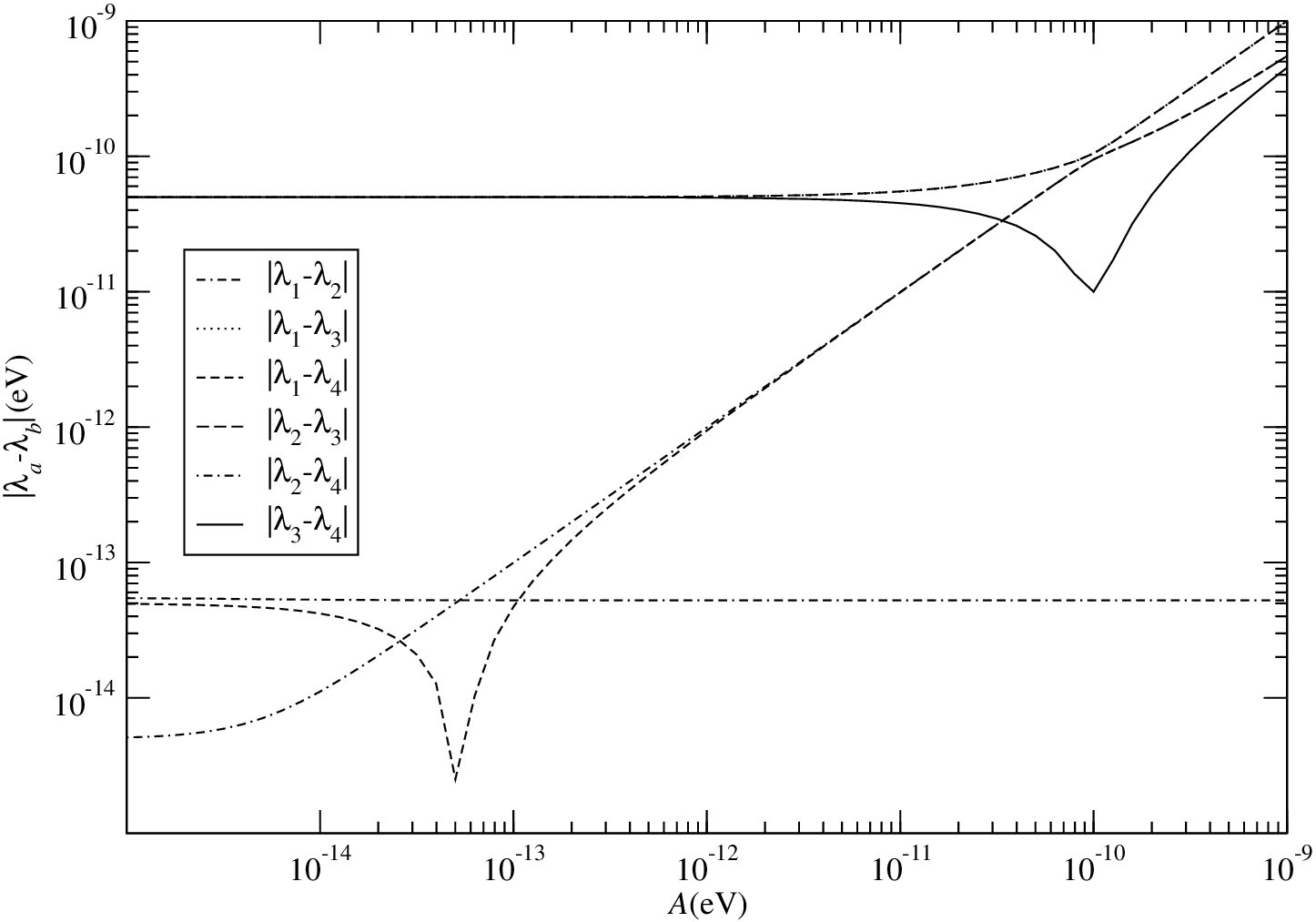}
  \end{tabular}
 \end{center}
 \caption{
 The difference $ |\lambda_{a} - \lambda_{b}|$ 
 $(a,b =1,2,3,4, ~a \neq b)$ as a function of the matter density $A$
 for a (3+1)-scheme, where 
 $ \theta_{12} = \frac{\pi}{4}$, $\theta_{23} = \frac{\pi}{4}$,
 $\theta_{13} = 0$, $\theta_{14} = \epsilon$,
 $\theta_{24} = \delta = 0$, $\theta_{34} = 0$,
 $\Delta m_{21}^2 \simeq 10^{-4} \mathrm{eV^2}$,
 $\Delta m_{32}^2 \simeq 10^{-3} \mathrm{eV^2}$, 
 $\Delta m_{41}^2 \simeq 1 \mathrm{eV^2}$,
 $E=10$ GeV
 and $\epsilon=0.1$.}
 \label{fig:A-lambda_31}
\end{figure}

As an illustration of the resonance phenomena, the energy differences
$|\lambda_a - \lambda_b| \quad (a,b=1,2,3,4, a \neq b)$
are plotted as a function of the matter density $A$ 
in Fig. \ref{fig:A-lambda_31}.
Here we assume that the electron number density $N_e$ is equal to
the neutral particle number density $N_n$ : $N_e = N_n$.
That is, $A_e=A$ and $A_n=\frac{1}{2}A$
in (\ref{eq:defA-e}) and (\ref{eq:defA-n}).
Note that $\lambda_a$'s are the effective neutrino energies in matter.
In Fig. \ref{fig:A-lambda_31}, 
the resonances occur when the energy levels in the presence of matter 
approach the values of each other.

\begin{figure}[htbp]
 \begin{center}
  \includegraphics[height=6cm,clip]{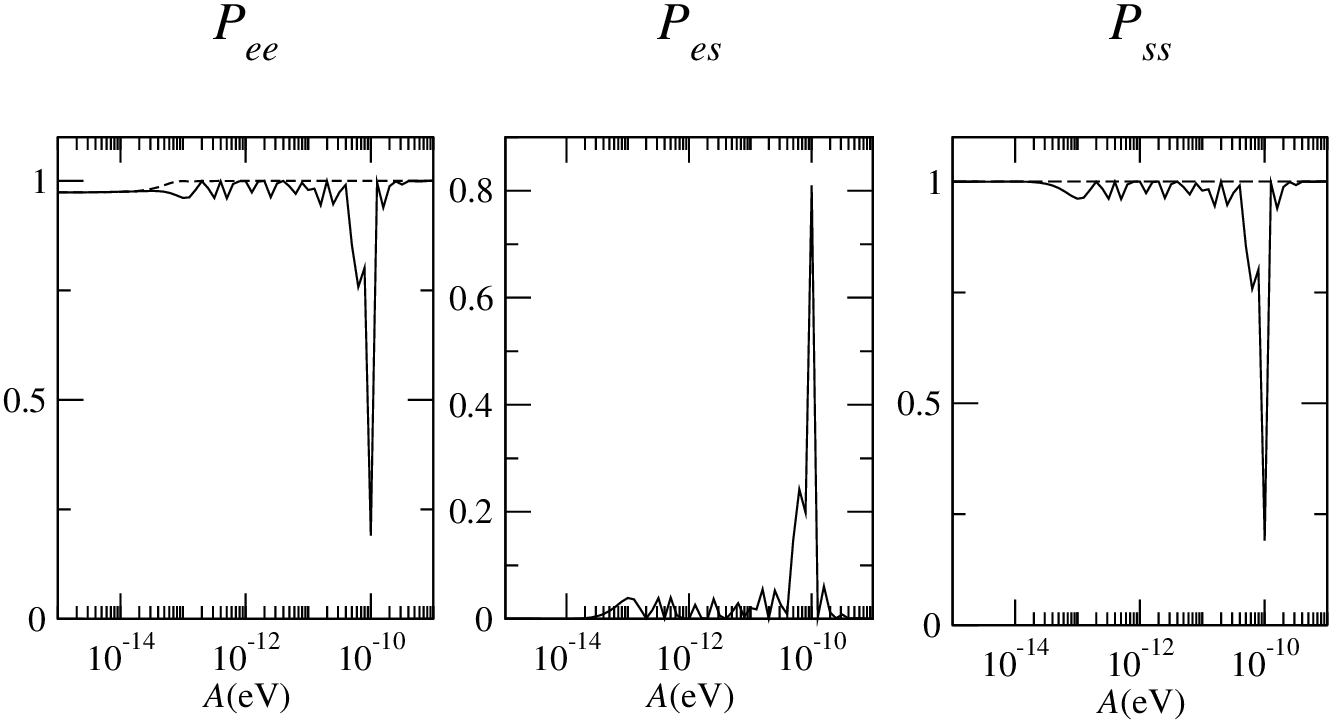}
 \end{center}
 \caption{
  The transition probabilities $P_{ee}$, $P_{es}$ and $P_{ss}$
 as a function of the matter density $A$
 for a (3+1)-scheme, where
 $ \theta_{12} = \frac{\pi}{4}$, $\theta_{23} = \frac{\pi}{4}$,
 $\theta_{13} = 0$, $\theta_{14} = \epsilon$,
 $\theta_{24} = \delta = 0$, $\theta_{34} = 0$,
 $\Delta m_{21}^2 \simeq 10^{-4} \mathrm{eV^2}$,
 $\Delta m_{32}^2 \simeq 10^{-3} \mathrm{eV^2}$, 
 $\Delta m_{41}^2 \simeq 1 \mathrm{eV^2}$,
 $E=10$ GeV and $L/E=6.46 \times 10^3 \mathrm{eV^{-2}}$.
 The solid and broken lines show the transition probabilities 
 for $\epsilon=0.1, 0$, respectively.}
 \label{fig:PeePesPss_31}
\end{figure}

The transition probabilities
(\ref{Eq:probability-of-transition}) for the neutrino oscillations
in matter as functions of the matter density~$A$
are shown in Fig.~\ref{fig:PeePesPss_31}.
They are some remarkable results showing the effects of the sterile neutrino.
Here, we set a parameter $\eta=1$.
The parameter $\eta$ is defined as 
$ L/E = \eta \times (2 R /{10 \mathrm{GeV}})= \eta \times 6.46 \times
10^3 \mathrm{eV^{-2}} $,
where $2R$ is the diameter of the earth :
  $R=3.23 \times 10^{13} \mathrm{eV^{-1}}  = 6378  \text{km}$.

From Fig.~\ref{fig:PeePesPss_31}, the following results can be derived.
If there is a little mixing of the neutrino \#1 and \#4
in the neutrino mass states, i.e., $\epsilon \neq 0$,
the fourth (sterile) neutrino effects appear as resonance.
For example, the probability $P_{ee}$ has a little transition 
beyond the matter density $A \sim 10^{-14}{\mathrm{eV}}$
and a sharp drop at $A \sim 10^{-10}{\mathrm{eV}}$.
\subsection{(2+2)-scheme}
The mass pattern about the (2+2)-scheme is 
shown by Fig.~\ref{fig:massscheme}(b).
We assume that both mass differences of the
$\nu_e$ and $\nu_s$ and of the $\nu_{\mu}$ and $\nu_{\tau}$ are small.   
The three kinds of the neutrino mass squared differences are put as follows,
\begin{eqnarray}
 \Delta m_{21}^2 &=& \Delta m^2_{\mathrm{LSND}}
  \simeq 1 \mathrm{eV^2}, \\
 \Delta m_{32}^2 &=& \Delta m^2_{\mathrm{atm}}
  \simeq 10^{-3} \mathrm{eV^2}, \\
 \Delta m_{41}^2 &=& \Delta m^2_{\mathrm{solar}}
  \simeq 10^{-4} \mathrm{eV^2}.
\end{eqnarray}
where the average $E$ of the neutrino energies is treated as $10 \mathrm{GeV}$.

The approximate mixing matrix for the (2+2)-scheme \cite{Barger00} is
\begin{equation}
 \left(
 \begin{array}{c}
  \nu_e \\
  \nu_{\mu} \\
  \nu_{\tau} \\
  \nu_s 
 \end{array}
 \right)
 =
\left(
\begin{array}{cccc}
 \frac{1}{\sqrt{2}}\cos \epsilon & \frac{1}{\sqrt{2}}\sin \epsilon &
  0 & \frac{1}{\sqrt{2}} \\
 -\frac{1}{\sqrt{2}}\sin \epsilon & \frac{1}{\sqrt{2}}\cos \epsilon &
  \frac{1}{\sqrt{2}} & 0 \\
 \frac{1}{\sqrt{2}}\sin \epsilon & -\frac{1}{\sqrt{2}}\cos \epsilon &
  \frac{1}{\sqrt{2}} & 0\\
 -\frac{1}{\sqrt{2}}\cos \epsilon & - \frac{1}{\sqrt{2}}\sin \epsilon&
  0 & \frac{1}{\sqrt{2}}\\
\end{array} 
\right)
 \left(
 \begin{array}{c}
  \nu_1 \\
  \nu_2 \\
  \nu_3 \\
  \nu_4 
 \end{array}
 \right),
 \label{Eq:22mixingmatrix}
\end{equation}
where $\epsilon$ is supposed to be small: $0 \leq \epsilon \leq 0.1 $.
These parameters resemble those in Ref.~\cite{Barger00},
except for $\theta_{13} = 0$.
The mixing matrix in (\ref{Eq:22mixingmatrix})
for the (2+2)-scheme is given from (\ref{eq:4x4mixingmatrix}) by taking 
\begin{equation}
 \theta_{14} = \frac{\pi}{4},\quad \theta_{23} = \frac{\pi}{4},\quad
  \theta_{12} = \epsilon,\quad \theta_{13} = 0,\quad
  \theta_{24} = 0,\quad \theta_{34} = 0
  \label{eq:angle22}.
\end{equation}
We take $\eta=1$, as shown in the (3+1) scheme.

\begin{figure}[htbp]
 \begin{center}
   \includegraphics[height=6cm,clip]{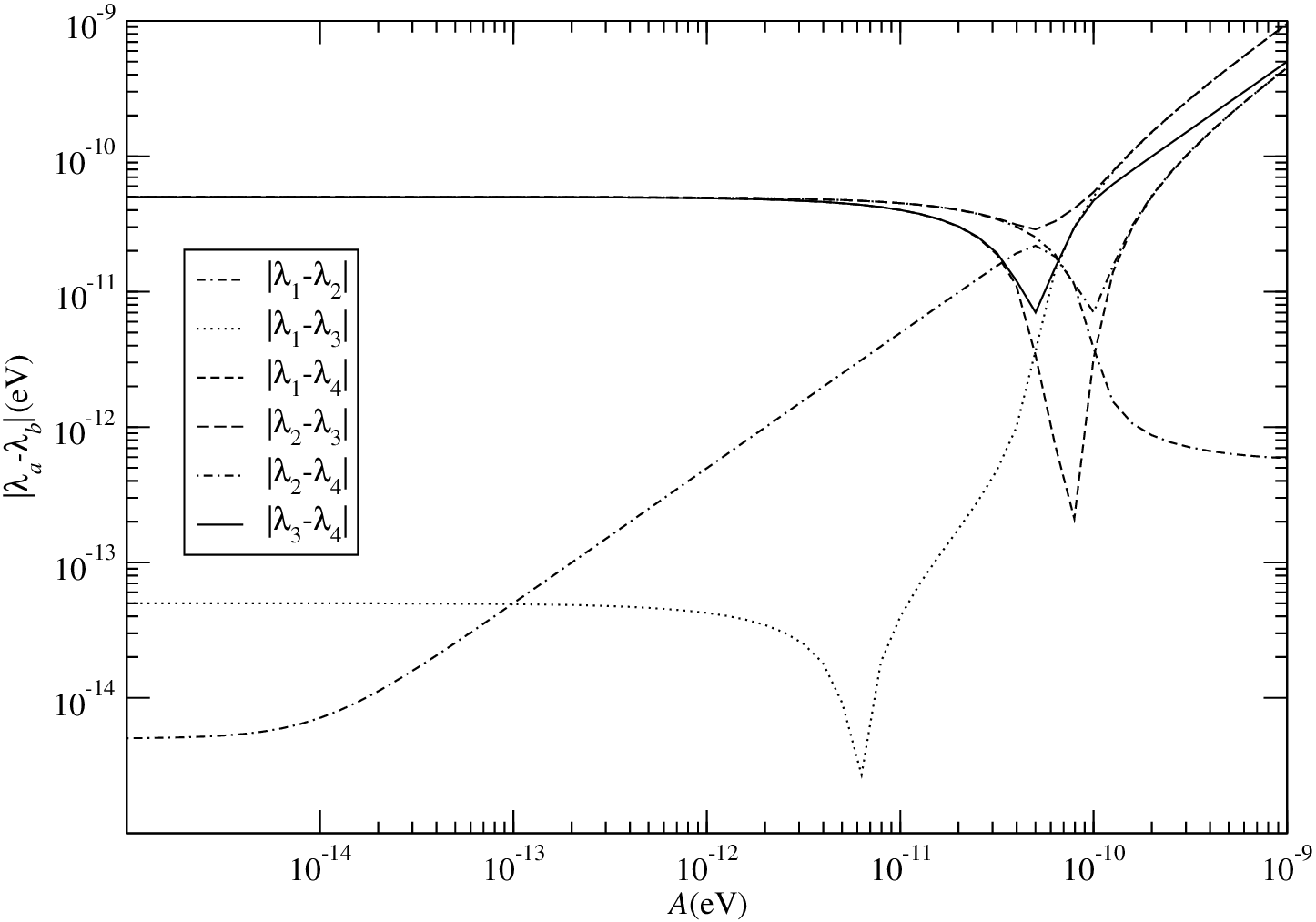}
 \end{center}
 \caption{
 The differences $|\lambda_a - \lambda_b| \quad (a,b=1,2,3,4)$
 as a function of the matter density $A$ for
 a (2+2)-scheme, where
 $\theta_{14} = \frac{\pi}{4}$, $\theta_{23} = \frac{\pi}{4}$,
 $\theta_{12} = \epsilon$, $\theta_{13} = 0$,
 $\theta_{24} = 0$, $\theta_{34} = 0$,
 $\Delta m_{21}^2 \simeq 1 \mathrm{eV^2}$,
 $\Delta m_{32}^2 \simeq 10^{-3} \mathrm{eV^2}$,
 $\Delta m_{41}^2 \simeq 10^{-4} \mathrm{eV^2}$,
 $E=10$ GeV and $\epsilon=0.1$.}
 \label{fig:A-lambda_22}
\end{figure}

\begin{figure}[htbp]
 \begin{center}
  \includegraphics[height=6cm,clip]{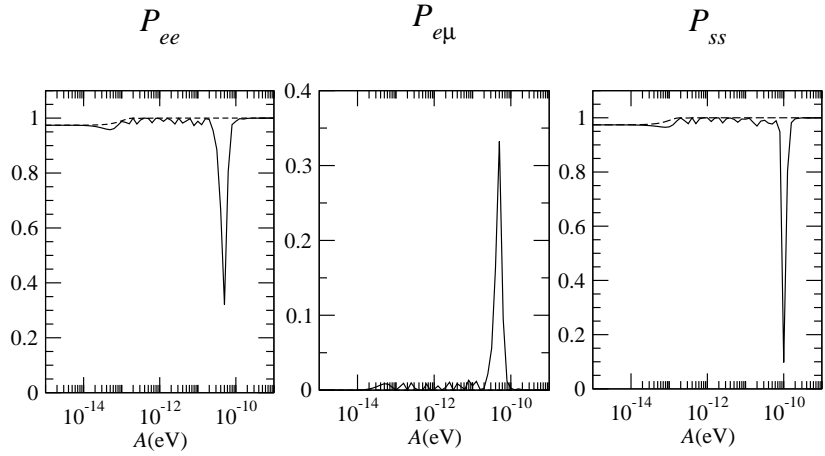}
 \end{center}
 \caption{
  The transition probabilities 
 $P_{ee}$, $P_{e\mu}$ and $P_{ss}$
 as a function of the matter density $A$ for
 a (2+2)-scheme, where
 $\theta_{14} = \frac{\pi}{4}$, $\theta_{23} = \frac{\pi}{4}$,
 $\theta_{12} = \epsilon$, $\theta_{13} = 0$,
 $\theta_{24} = 0$, $\theta_{34} = 0$,
 $\Delta m_{21}^2 \simeq 1 \mathrm{eV^2}$,
 $\Delta m_{32}^2 \simeq 10^{-3} \mathrm{eV^2}$,
 $\Delta m_{41}^2 \simeq 10^{-4} \mathrm{eV^2}$,
 $E=10$ GeV and  $L/E=6.46 \times 10^3 \mathrm{eV^{-2}}$.
 The solid and broken lines show the transition probabilities 
 for $\epsilon=0.1, 0$, respectively.
 }
 \label{fig:PeePemPss_22}
\end{figure}

In the (2+2)-scheme, the results of the energy differences
$|\lambda_a - \lambda_b|$ $(a,b=1,2,3,4)$ are 
presented in Fig.~\ref{fig:A-lambda_22} 
as a function of the matter density $A$.
The transition probabilities $P_{\alpha\beta}$
for the neutrino oscillations in matter 
are shown in Fig.~\ref{fig:PeePemPss_22}.

We assume that two mixings between $\nu_e$ and $\nu_s$ and 
between $\nu_{\mu}$ and $\nu_{\tau}$ are maximal,
and the other four mixings are minimal as (\ref{eq:angle22}).
Some transition probabilities $P_{\alpha\beta}$ are shown
in Fig.~\ref{fig:PeePemPss_22}, and $P_{e\tau}$ resembles
$P_{e\mu}$ in a probability pattern.
The transition between two neutrinos,
of which masses are clearly distinguished each other,
occurs beyond $A \sim 5 \times 10^{-11}\mathrm{eV}$.

\section{Discussion}\label{chap:discussion}
The main result of our analysis is given by the time evolution operator
(\ref{eq:flavor-t-evol}) for the four-neutrinos in matter.
The time evolution operator (\ref{eq:flavor-t-evol})
in the flavor eigenstate is expressed as a finite sum of
elementary functions in the matrix elements of the Hamiltonian
(\ref{eq:HamiltonianInMatter}).
The transition probabilities in matter have been 
given by (\ref{Eq:probability-of-transition}).
We also have analyzed the matter effects of the transition probabilities,
assuming that there are four kinds of neutrinos and
that the fourth (sterile) neutrino has
a little mixing with the other neutrinos.

The resonance between an active neutrino
($\nu_{e}$, $\nu_{\mu}$ or $\nu_{\tau}$) and the sterile neutrino $\nu_{s}$
occurs at the matter density $A \simeq 10^{-10} \mathrm{eV}$. 
Is this matter density realistic?
We consider about the density of the sun
(see Appendix~\ref{chap:electron-density}).
Using a solar model\cite{Bahcall89}, 
the electron matter density $A$ in the sun is 
$1.06 \times 10^{-11} \mathrm{eV}$ at the center of the sun and
$2.78 \times 10^{-16} \mathrm{eV}$ at the surface of the sun, respectively.
The average of the electron matter density in the sun
is $1.40 \times 10^{-13}\mathrm{eV}$.
Thus, the electron matter density in the sun has
the values from $10^{-16}\mathrm{eV}$ to $10^{-11}\mathrm{eV}$.
Our result $A \simeq 10^{-10}\mathrm{eV}$,
at which the four-neutrino resonance occurs, is very large.
Therefore, one may not observe the sterile neutrino resonance realistically,
even if it exists.
To detect the fourth neutrino resonance,
one needs matter of which density is about $10^{-10}\mathrm{eV}$. 

The average of the electron matter density in the earth is
about $4.92 \times 10^{-13} \mathrm{eV}$ 
(see Appendix \ref{chap:electron-density}).
This is the same order as the matter density of the sun.
So it will be difficult to find the resonance of the sterile neutrino
near the earth.

\begin{figure}[htbp]
 \begin{center}
  \includegraphics[height=6cm,clip]{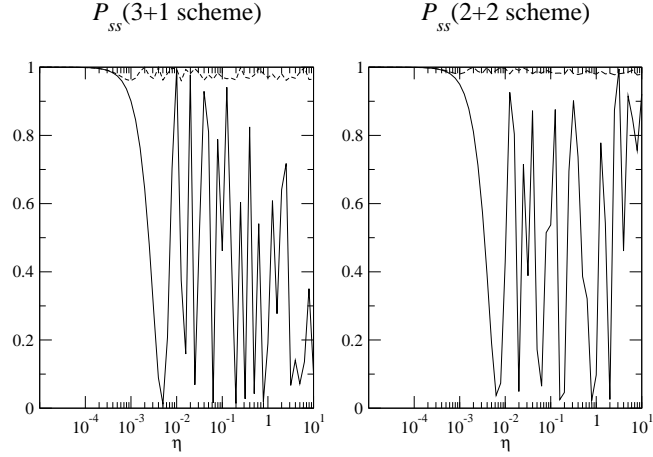}
 \end{center}
 \caption{
 The surviving probabilities $P_{ss}$ of the sterile neutrino transition
 as a function
 of the rate $\eta$ for a (3+1)-scheme and a (2+2)-scheme,
 where $L/E = \eta \times 6.46 \times 10^3 {\mathrm{eV^{-2}}}$.
 We compare $P_{ss}$ at two kinds of the matter densities for
 $A=1.40 \times 10^{-13}$[eV](the broken line) and 
 $A=10^{-10}$[eV](the solid line). 
 }
 \label{fig:Pss-eta}
\end{figure}

Figure~\ref{fig:Pss-eta} shows the transition probability $P_{ss}$,
which is the sterile neutrino surviving probability,
as a function of the parameter $\eta$ 
for the (3+1)-scheme and the (2+2)-scheme.
From Fig.~\ref{fig:Pss-eta}, 
the sterile neutrino transition occurs for $\eta > 10^{-4}$.

The transition from an active neutrino to a sterile one may be observed
if neutrino passes through the matter which has very high density.
In the same condition, the reverse transition may occur.
The above argument about the transition probability 
for four-neutrino oscillations 
tells that we may not realistically see the sterile neutrino resonance,
even if the sterile neutrino exists.

\acknowledgments
The authors are grateful to Professor T. Maekawa for valuable suggestions.

\appendix
\section{Electron number density in the natural system of units}
\label{chap:electron-density}
In this paper, we refer to the Ref.~\cite{Caso98} for the physical constants.
The electron number density $N_e$ is connected with the matter density $A$
by (\ref{eq:defA-e}).
Using the Fermi weak coupling constant
$G_F / {(\hbar c)}^3 = 1.17 \times 10^{-5} [{\text{GeV}}^{-2}]$,
\begin{equation}
 A [\text{eV}]= \sqrt{2} G_F N_e 
  = 1.27 \times 10^{-43}[\text{eV} \cdot {\text{m}}^3]
  \cdot N_e[1/{\text{m}}^3],\label{eq:Ne2A}
\end{equation}
where $\hbar c = 197 [\text{MeV fm}]$.

We discuss the matter density in the sun, using a solar model \cite{Bahcall89}.
From the model, the electron number density $N_e$ depends on 
the distance $R$ from the center of the sun :
\begin{equation}
 N_e(x) = 98.19 ~ N_A e^{- 10.55 x}~[{\mathrm {cm^{-3}}}], \quad
  x \equiv \frac{R}{R_{\mathrm{sun}}},
\end{equation}
where $N_A=6.02 \times 10^{23}$ and $R_{\textrm{sun}}$ are
the Avogadro constant and the radius of the sun, respectively.

In Table \ref{table:electron-density},
the electron matter densities of the sun and the earth are listed.
And we assume that the ratio of the electron mass to the nucleon mass 
is 1 : 2000, where the mass of electron is $9.11 \times 10^{-31}$[kg].
From Table \ref{table:electron-density},
the electron number density in the sun has the values
from $10^{-16}\mathrm{eV}$ to $10^{-11}\mathrm{eV}$.  

\newpage
%
\begin{table}[htbp]
 \begin{center}
  \begin{tabular}{|l|c|c|c|}
   \hline
   & $A$[eV] & $N_e$ [1/{cm$^3$}] &
   $\rho$ [g/{cm$^3$}]\\
   \hline
   $R/R_{\mathrm{sun}}=0$ & 1.06 $\times 10^{-11}$ & 5.91 $\times 10^{25}$ 
   & 108\\
   $R/R_{\mathrm{sun}}=0.41$ & 1.40 $\times 10^{-13}$& 7.82 $\times 10^{23}$
   & 1.42\\
   $R/R_{\mathrm{sun}}=1$ & 2.78 $\times 10^{-16}$& 1.55 $\times 10^{21}$
   & 2.82 $\times 10^{-2}$ \\
   \hline
   the earth & 5.44 $\times 10^{-13}$
   & 3.03 $\times 10^{24}$ & 5.52 \\ \hline
  \end{tabular}
 \end{center}
 \caption{
 The electron matter densities of the sun and the earth.
 The value of $A$ at $x={R}/{R_{\mathrm{sun}}}=0.41$
 corresponds to the average density in the sun.
 The mass density $\rho$ is shown as $\rho = 2000 m_e N_e$,
 where $m_e$ is the electron mass, and 
 we put the ratio of the electron mass to the nucleon mass is 1 : 2000.
 We assume that the density in the earth is $5.52 {\mathrm{g/cm^3}}$.}
 \label{table:electron-density}
\end{table}

\begin{thebibliography}{99}
 \bibitem{Ahmad01}Q. R. Ahmad et al., 
	 Phys. Rev. Lett. \textbf{87}, 071301 (2001).
 \bibitem{Cleveland98}B. T. Cleveland et al.,
	 Astrophys. J. \textbf{496}, 505 (1998).
 \bibitem{Hampel99}W. Hampel et al.,
	 Phys. Lett. B \textbf{447}, 127 (1999).
 \bibitem{Fukuda96}Y. Fukuda et al.,
	 Phys. Rev. Lett. \textbf{77}, 1683 (1996).
 \bibitem{Fukuda99}Y. Fukuda et al.,
	 Phys. Rev. Lett. \textbf{82}, 1810 (1999).
 \bibitem{Scholberg99}K. Scholberg, 
	 for the Super-Kamiokande Collaboration,
	 presented at 8th International Workshop on Neutrino Telescopes, 
	 Venice, 1999, e-print hep-ex/9905016.
 \bibitem{Abdurashitov99}J. N. Abdurashitov et al.,
	 Phys. Rev. C \textbf{60}, 055801 (1999).
 \bibitem{Athanassopoulos96}C. Athanassopoulos et al.,
	 Phys. Rev. Lett. \textbf{77}, 3082 (1996).
 \bibitem{MSW85} S. P. Mikheyev and A. Y. Smirnov,
	 Yad. Fiz. \textbf{42}, 1441 (1985) [Sov. J. Nucl. Phys. \textbf{42},
	 913 (1985)];
	 S. P. Mikheyev and A. Y. Smirnov, Nuovo Cimento C
	 \textbf{9}, 17 (1986); L. Wolfenstein, Phys. Rev. D \textbf{17}, 2369
	 (1978);
	 L. Wolfenstein, {\it ibid.} \textbf{20}, 2634 (1979).
 \bibitem{Bilenky98}S. M. Bilenky, C. Giunti and W. Grimus,
         Prog. Part. Nucl. Phys. \textbf{43}, 1 (1999).
 \bibitem{MNS62}Z. Maki, M. Nakagawa and S. Sakata,
	 Prog. Theor. Phys. \textbf{28}, 870 (1962).
 \bibitem{Ohlsson00}T. Ohlsson, H. Snellman,
         J. Math. Phys. \textbf{41}, 2768 (2000),
	 {\it ibid.} \textbf{42}, 2345 (2001).
 \bibitem{Barger99}V. Barger, Y. B. Dai, K. Whisnant and B. L. Young,
	 Phys. Rev. D \textbf{59}, 113010 (1999).
 \bibitem{Barger00}V. Barger, B. Kayser, J. Learned, 
	 T. Weiler and  K. Whisnant,
        Phys. Lett. B \textbf{489}, 345 (2000).
 \bibitem{Spiegel68}
	 {\it Mathematical Handbook of Formulas and Tables}, 
	 edited by M. R. Spiegel (McGraw-Hill, Inc., New York, 1968).
 \bibitem{Caso98}Particle Data Group, C. Caso et al.,
	 Review of Particle Physics, Eur. Phys. J. C \textbf{3}, 1 (1998).
 \bibitem{Bahcall89}J. N. Bahcall, {\it Neutrino Astrophysics}, 
	 (Cambridge Univ. Press, New York, 1989) p. 101.
\end{thebibliography}
\end{document}